\documentclass[useAMS,usenatbib]{mn2e}
\usepackage{graphicx}
\usepackage{txfonts}
\usepackage{natbib}

\newcommand{\kms}    {km~s$^{-1}$}
\def\apj{ApJ}%
\def\apjl{ApJ}%
\def\apjs{ApJS}%
\def\aap{A\&A}%
\def\mnras{MNRAS}%
\title[Methanol maser polarization in W3(OH)]{Methanol Maser Polarization in W3(OH)}
\author[W. H. T. Vlemmings, L. Harvey-Smith and R. J. Cohen]{W. H. T. Vlemmings$^{1}$\thanks{E-mail: wouter@jb.man.c.uk}, L. Harvey-Smith$^{2}$ and R. J. Cohen$^{1}$\\
$^{1}$Jodrell Bank Observatory, University of Manchester, Macclesfield, Cheshire, Sk11 9DL, England\\
$^{2}$Joint Institute for VLBI in Europe, Postbus 2, 7990~AA Dwingeloo, The Netherlands}
\begin{document}

\date{Received 2006, May 24. Accepted 2006, June 9.}

\pagerange{\pageref{firstpage}--\pageref{lastpage}} \pubyear{2002}

\maketitle

\label{firstpage}

\begin{abstract}
 We present the first 6.7~GHz methanol maser linear polarization map
 of the extended filamentary maser structure around the compact
 {\sc Hii} region W3(OH). The methanol masers show linear polarization
 up to $\sim 8$~per~cent and the polarization angles indicate a
 magnetic field direction along the North-South maser
 structure. The polarization angles are consistent with those measured
 for the OH masers, taking into account external Faraday rotation
 toward W3(OH), and confirm that the OH and methanol masers are found
 in similar physical conditions. Additionally we discuss the Zeeman
 splitting of the 6.7~GHz methanol transition and present an upper
 limit of $\sim 22$~mG for the magnetic field strength in the maser
 region.  The upper limit is fully consistent with the field strengths
 derived from OH maser Zeeman splitting.
\end{abstract}


\section{Introduction}

Through polarization observations, maser are unique probes of the
magnetic fields that are thought to play an important role in
star-forming regions (SFRs). Observations of the Zeeman splitting of
several different transitions of the OH masers indicate mG level
magnetic fields, and the linear polarization of the masers has also
been used to determine the structure of the magnetic field
\citep[e.g.][]{Etoka05, Bartkiewicz05}. In addition to the OH masers,
the H$_2$O masers at 22~GHz have also been used to determine the
magnetic field strength and structure in the more dense areas of SFRs
\citep[e.g.][]{Sarma01, Imai03, Vlemmings06}. These observations have
revealed magnetic field strengths of several tens to hundreds of mG.

The 6.7~GHz methanol maser is a strong tracer of high-mass
star-formation \citep{Menten91} and is typically one of the brightest
masers in those regions. The masers occur in regions with similar
densities and temperatures to the ground state OH masers, where
enhanced densities of both molecular species are to be expected as a
result of material from grain mantles being evaporated under the
influence of weak shocks \citep{Hartquist95}. Theoretical modelling
has shown that both masers can be pumped simultaneously
\citep{Cragg02}. Similar to H$_2$O, methanol is a non-paramagnetic
molecule. As a result, the Zeeman splitting is only a small fraction
of the typical maser line width. Thus, the determination of the
magnetic field strength using 6.7~GHz methanol masers requires
sensitive and high spectral resolution observations. As yet, no successful
methanol Zeeman experiment has been reported. Only recently have
the first measurements of linear polarization of the 6.7~GHz methanol
maser been made. Observations by \citet{Ellingsen02} indicate a
fractional linear polarization up to $10$~per~cent, although no
polarization maps have been produced until now.

In this letter the polarization properties of the 6.7~GHz methanol
masers in the massive star-forming region W3(OH) are presented. The
polarization has been determined using MERLIN\footnote{MERLIN is a
  national facility operated by the University of Manchester at
  Jodrell Bank Observatory on behalf of PPARC} observations presented
by \citet[][~hereafter HSC06]{HarveySmith06}, where the focus has been
on the astrometry and distribution of the methanol masers.

W3(OH) is one of the most intensively studied star-forming regions and
contains one of the most luminous ultracompact {\sc Hii} (UC{\sc Hii})
regions. It is located in the Perseus spiral arm of the Galaxy at a
distance of $1.95\pm0.04$~kpc as determined using maser astrometry
\citep{Xu06}. It is the site of several young high- and
intermediate-mass stars, the most prominent of which is the UC{\sc
  Hii} region itself, which is ionized by an embedded zero-age
main-sequence O-star \citep[e.g][]{Dreher81}. The W3(OH) region
displays strong emission from a wide variety of maser species
\citep[e.g][~and references therein]{Wright04a, Etoka05}.  Strong OH
and methanol masers are seen projected onto the UC{\sc Hii} region and
strong H$_2$O masers have been found towards the related Turner-Welch
(TW) Object \citep{Turner84}.

\section{Observations}

The 6.7~GHz methanol masers of the UC{\sc Hii} region W3(OH)
were observed on December 10-11 in 2004 using MERLIN.  A 500~kHz
bandwidth with 512 spectral channels was used, centered at $V_{\rm
  LSR}=-45$~\kms, resulting in a velocity resolution of
0.044~\kms\ and a total velocity coverage of 22.5~\kms. The
observations have been discussed in detail in HSC06.

The initial calibration steps were as described in HSC06, but a
separate calibration for optimal polarization measurements was carried
out. Instrumental feed polarization was determined using the AIPS task
PCAL. After correction for the R-L phase offset, self-calibration was
performed on a strong, isolated maser feature using only the right
circular polarization. The calibration results were then applied to
both polarizations. 3C286 was used to calibrate the polarization
angle. Finally, image cubes were created in Stokes I, Q, U and V, as
well as in right and left circular polarization (RHC and LHC), by
mapping with a circular restoring beam of 50~mas.

Since the self-calibration for the polarization experiment was
performed on the RHC data alone, not all the flux found in HSC06 was
recovered and the rms noise level is somewhat increased. The rms noise
in the Stokes I, Q, U, V maps and the RHC and LHC maps was $\sim
15$~mJy~beam$^{-1}$ and $\sim 20$~mJy~beam$^{-1}$ respectively, in
channels that were free of bright maser features. However, in the
brightest channels, the rms noise levels were significantly higher due
to dynamic range effects. Consequently, only the polarization of
several of the brightest ($>70$~Jy) maser features could be
determined.

\begin{table}
 \centering
  \caption{Linearly polarized CH$_3$OH maser features}
  \begin{tabular}{ccccc}
  \hline
 Maser feature & $P_L$ & $\chi$ & R.A. (J2000) & Dec (J2000) \\
 nr.$^a$ & (per cent) & $(^\circ)$ & $(02^h27^m~~^s$ & $(61^\circ52'~~'')$ \\
 \hline
5 & $0.2 \pm 0.03$ & $-70 \pm 6$ & 3.7994 & 23.483 \\
10 & $1.9 \pm 0.3$ & $-71 \pm 6$ & 3.8113 & 24.091 \\
18 & $1.4 \pm 0.4$ & $-94 \pm 3$ & 3.8142 & 23.942 \\
19 & $2.4 \pm 0.4$ & $-55 \pm 6$ & 3.7947 & 24.560 \\
21$^b$ & $3.2 \pm 0.5$ & $-62 \pm 2$ & 3.8199 & 25.231 \\
25$^b$ & $3.3 \pm 0.4$ & $-2 \pm 1$ & 3.8175 & 25.252 \\
27 & $1.2 \pm 0.3$ & $-78 \pm 3$ & 3.7039 & 25.333 \\
\hline
\multicolumn{5}{l}{$^a$ from HSC06} \\
\multicolumn{5}{l}{$^b$ features that are part of the broadline region} \\
\multicolumn{5}{l}{~~~ which displays large variations in $P_L$ and $\chi$ (see text)} \\
\end{tabular}
\label{Table:res}
\end{table}

\begin{figure}
\includegraphics[width=8.0cm]{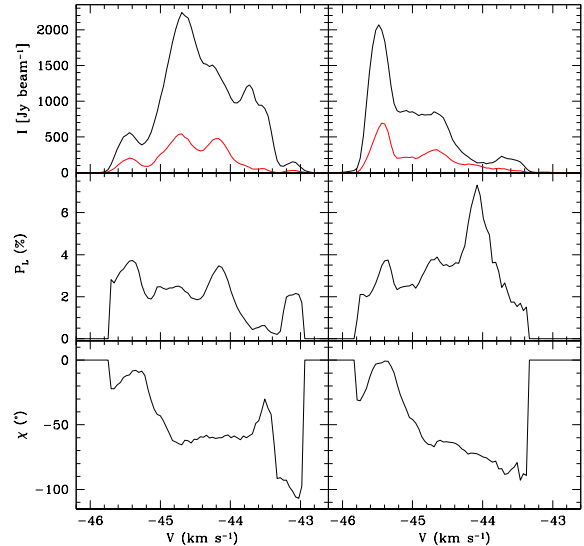}
 \caption{Spectra, fractional linear polarization $P_L$ and
   polarization angle $\chi$ for two of the maser features in the
   bright broadline maser region. The left figure is feature $21$ and
   the right figure is feature $25$. The top panels are the total
   intensity flux (black) and the polarized intensity flux (red),
   where the polarized intensity has been multiplied by a factor of
   10. The middle panels show $P_L$ and the lower panels show the
   variation of $\chi$ across the complex maser spectra. }
 \label{Fig:spec}
\end{figure}

\section{Results}

HSC06 show that the peak flux densities of the 6.7~GHz methanol masers
of W3(OH) range from $0.28$ to $2242$~Jy~beam$^{-1}$. We have detected
linear polarization on 7 of the strongest maser features with peak
flux densities $>70$~Jy. For these features, the positions from HSC06,
the fractional linear polarization $P_L$ and the polarization angle
$\chi$ are given in Table.~\ref{Table:res}. Here the maser features
are denoted with the numbers assigned in HSC06 and both $P_L$ and
$\chi$ are flux weighted averages across the maser feature. The errors
are the associated $1\sigma$ errors.

As discussed in HSC06, 80~per~cent of the 6.7~GHz maser flux
originates from a compact region containing several maser
  features with $V_{\rm LSR}$ from $-47.5$ to $-42.6$~\kms. As this
  region is characterised by masers with a broad line profile, the
  region will be called the 'broadline' region, in correspondence with
  the terminology used in HSC06.
The strongest linear polarization is also found in this region, with
values of $P_L$ between $\sim 1$~per~cent and $\sim 8$~per~cent.
Fig.~\ref{Fig:spec} shows the spectra, $P_L$ and $\chi$ for features
$21$ and $25$.  It can also be seen that $\chi$ changes smoothly
across the spectrum, from $\chi\approx 0^\circ$ for the masers at
$V_{\rm LSR}\approx -45.4$~\kms\ to $\chi\approx-110^\circ$ for the
masers at $V_{\rm LSR}\approx -43.0$~\kms.

\begin{figure*}
\includegraphics[width=14.0cm]{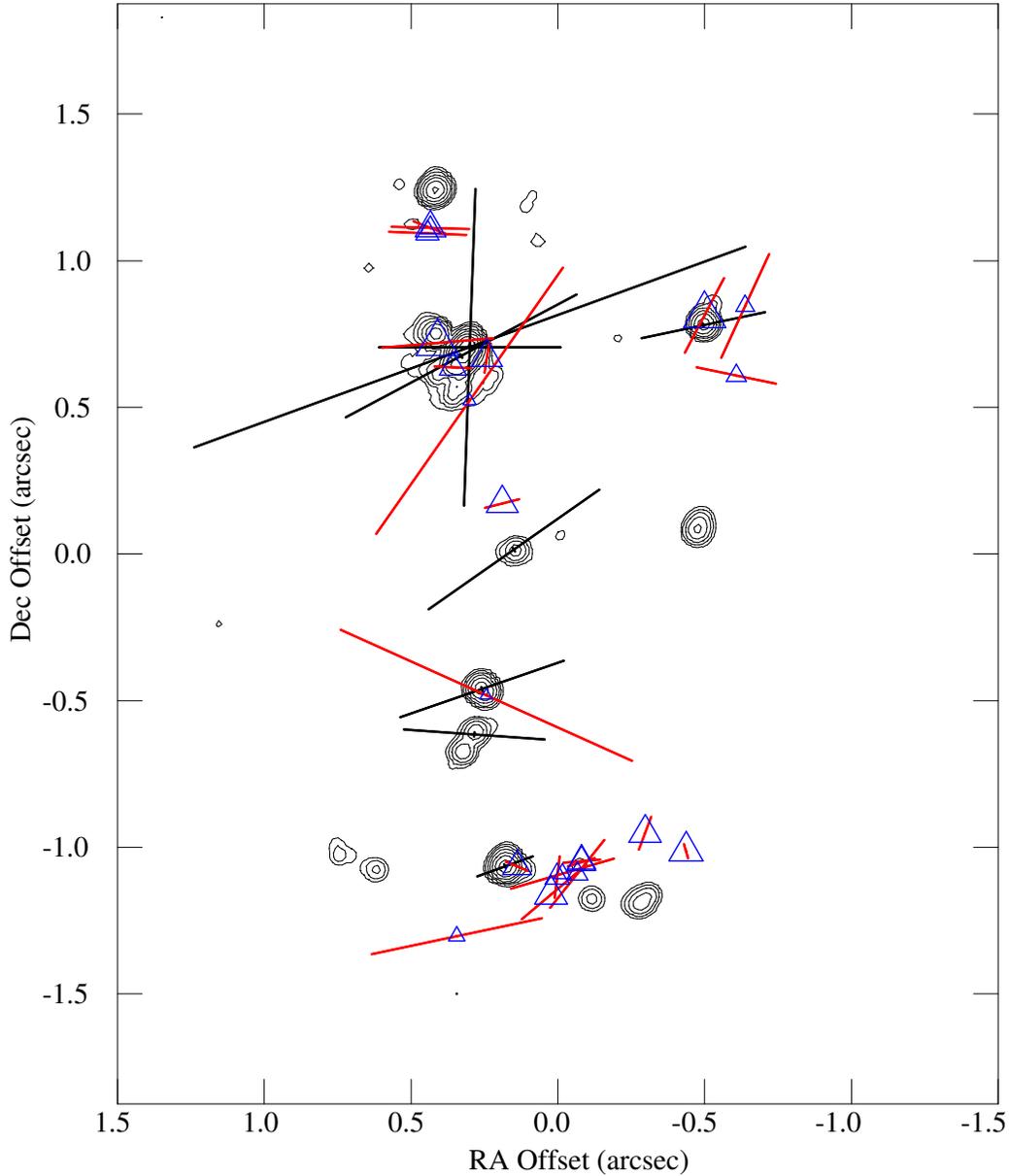}
 \caption{The methanol masers of W3(OH) (contours) including the
   polarization vectors (black) scaled linearly according the
   fractional linear polarization $(P_L)$. The positions in the map
   are indicated with respect to R.A.(J200)$=02^h27^m03.7743^s$,
   Dec(J2000)$=61^\circ52'24.549''$. The broadline maser region is located at an offset of (0.32,0.69) arcsec. The blue triangles denote the
   main line OH masers from \citet{Wright04b} for which polarized
   intensity was detected at $5\sigma$ significance (polarized flux
   $>75$~mJy), and the red vectors are their linearly scaled
   polarization vectors. The main line OH maser polarization vectors
   lengths are scaled down by a factor of 5 with respect to the
   lengths of the methanol maser polarization vectors.}
 \label{Fig:map}
\end{figure*}

Excluding the polarization angle gradient across the broadline
region, the other maser features show a fairly constant polarization
angle. A map of the strongest 6.7~GHz maser features together with their
polarization angles is shown in Fig.~\ref{Fig:map}. 

We also produced a circular polarization $V$ image cube. However, both
because of dynamical range effects and heavy spectral blending of
the various maser features we were unable to detect any significant
circular polarization above a level of $\sim 2$~per~cent. \citet{Modjaz05}
describe a cross-correlation method to determine Zeeman splitting,
which is specifically suited for blended maser line of
non-paramagnetic maser species, assuming the magnetic field is mostly
constant across the spectrum. This method has been used to determine the
upper limit of the magnetic field strength from the 22~GHz H$_2$O
maser polarization spectra of NGC~4258, and it was shown that the
uncertainty of the method, which uses RHC and LHC polarization
spectra, is similar to that of the standard method using circular polarization
measurements. We thus produced separate image cubes in RHC and LHC
polarizations and determined the RHC and LHC polarized spectra of the
brightest maser features. Unfortunately, we were still limited by the
dynamic range in the RHC and LHC polarization spectra and were only
able to determine an upper limit to the Zeeman splitting on the
brightest maser features of the broadline region. This upper
limit to the Zeeman splitting of the 6.7~GHz methanol masers of W3(OH)
is $\Delta V_Z < 1.1\times10^{-3}$~\kms.

\section{Discussion}

\subsection{Linear Polarization}

\subsubsection{Fractional Polarization}

The fractional linear polarization of the 6.7~GHz methanol masers is
comparable to that found in the Southern star-forming region,
NGC6334F, by \citet{Ellingsen02}. For the masers in W3(OH), the error
weighted median $\langle P_L\rangle_{\rm med}=1.8\pm0.7$~per~cent,
while the maximum linear polarization fraction is $\sim
8$~per~cent. From a small sample of methanol maser sources,
\citet{Ellingsen02} has concluded that polarization of 6.7~GHz maser
is more widespread than of 12.2~GHz methanol maser, consistent with
6.7~GHz masers being more saturated. However, \citet{Ellingsen02} has
also found that the polarization characteristics of both methanol
maser transitions in NGC6334F are strikingly similar when comparing
6.7~GHz observations with observations at 12.2~GHz by
\citet{Koo88}. Polarization of the 12.2~GHz methanol masers in W3(OH)
has also been observed by \citet{Koo88}, who find a linear
polarization of $\sim 4$~per~cent. Thus from our observations, we
conclude that also for the methanol masers in W3(OH), where the
12.2~GHz maser velocity range overlaps exactly with the 6.7~GHz maser
velocity range, the polarization characteristics of both transitions
are very similar.

\subsubsection{Polarization Angles \& Faraday Rotation}

We have for the first time mapped the linear polarization of 6.7~GHz
methanol masers. We find that, as seen in Fig.~\ref{Fig:map}, the
polarization angles in W3(OH) are similar across the entire maser region with
the exception of those in part of the broadline region. The error
weighted median of the polarization $\langle\chi\rangle_{\rm med}=-67
\pm 9^\circ$.

As shown in Fig.~\ref{Fig:spec}, the broadline region displays a
smooth but considerable polarization angle gradient. High spatial resolution Very Long Baseline Array (VLBA)
polarization observations have been made of the 13.44~GHz OH masers
that are located within the broadline methanol maser region. These
indicate that there is also a significant spread in the polarization
angles of those masers (Diamond et al., in preparation).

In Fig.~\ref{Fig:map} we have also indicated the polarization vectors
of the ground-state OH masers at 1.6~GHz from the electronic tables
of \citet{Wright04b}, for which linear polarization has been detected at
more than a $5\sigma$ level. The OH maser polarization angles show a
wider spread, with the median polarization angle
$\langle\chi\rangle_{\rm med_{OH}}=104 \pm 27^\circ$, which, for the 1.6~GHz OH masers, has been
attributed to Faraday rotation along the maser path
\citep[e.g][]{Fish06}.  Faraday rotation can be described by:
\begin{equation}
\Phi{\rm [^\circ]} = 4.17\times10^6~D{\rm [kpc]}~n_e{\rm [cm^{-3}]}~B_{\rm ||}{\rm [mG]}~\nu^{-2}{\rm [GHz]},
\end{equation}
where $D$ is the length of the path over which the Faraday rotation
occurs , $n_e$ and $B_{\rm ||}$ are respectively the average electron
density and magnetic field along this path and $\nu$ is the
frequency. If the Faraday rotation is significant ($\Phi > 1$~rad)
along the maser amplification path, but not too large ($\Phi < 1$~rad)
along a single gain length (typically $<1/20$th of the amplification
path), the Faraday rotation will not completely inhibit the linear
polarization, but will still cause a significant spread of
polarization angles for different maser features. Since Faraday
rotation scales with $\nu^{-2}$, the higher frequency methanol masers
are mostly unaffected by internal Faraday rotation, as is
illustrated by the difference in the spread of the polarization
angles of the OH and methanol masers.

In addition to the internal Faraday rotation, external Faraday
rotation makes the relationship between the polarization angle and
magnetic field direction uncertain for the lower frequency
observations. Furthemore, it makes a direct comparison between the
1.6~GHz OH maser and the 6.7~GHz methanol maser polarization angles
difficult. For typical values of the interstellar electron density
$n_e\approx 0.04$~cm$^{-3}$ and magnetic field $B_{\rm ||}\approx
1.5~\mu$G, and assuming a distance to W3(OH) of $D\approx 1.95$~kpc,
the Faraday rotation at 6.7~GHz is $\Phi\approx 11^\circ$. The Faraday
rotation at 1.6~GHz on the other hand is $\Phi\approx
190^\circ$. Thus, the difference between the median polarization
angles at 6.7~GHz and 1.6~GHz, $\Delta\chi=171\pm28^\circ$, is probably
solely caused by external Faraday rotation. 

An additional complication in determining the direction of the
magnetic field from the linear polarization observations is the
$90^\circ$ ambiguity in the relationship between the polarization
angle and the magnetic field direction \citep{Goldreich73}. As the
linear polarization of the OH masers probably arises from the
elliptical polarization of the $\sigma$-components, the polarization
angle of the OH masers is perpendicular to the magnetic field. Thus,
since the difference in polarization angle between the 1.6~GHz OH
masers and the 6.7~GHz methanol masers can be fully explained by
external Faraday rotation, it would appear that the 6.7~GHz
polarization angles are also perpendicular to the magnetic
field. Correcting the methanol maser polarization angles for the
external Faraday rotation yields a position angle of $\sim
10^\circ$--$15^\circ$ for the magnetic field in the methanol maser
structure of W3(OH), which is nearly parallel to the extended 1.6 arcsecond long N-S maser filament, described in HSC06. This is consistent with the model by \citet{Dodson04}, in which linear methanol sources trace planar shocks propagating nearly perpendicular to the line-of-sight. According to this model the magnetic field should be parallel to the elongated (shock) structure.

\subsection{Magnetic field strength}

The methanol molecule is a non-paramagnetic molecule and as a result
the Zeeman splitting under the influence of a magnetic field is
extremely small. The split energy, $\Delta E_Z$, of an energy level
under the influence of a magnetic field, $B$, can be described as
$\Delta E_z = 10^4\times g \mu_N M_J B$, where $M_J$ denotes the
magnetic quantum number for the rotational transition described with
the rotational quantum number $J$, $B$ is the magnetic field strength
in Gauss 
, $\mu_N$ is the nuclear magneton and $g$
is the Land\'e $g$-factor. The Zeeman effect is determined by the
Land\'e $g$-factor, which for methanol has been investigated many
years ago by \citet{Jen51}, who found empirically that it is probably
an average of the true $g$-factor of several interacting states and
can be described by the equation:-
\begin{equation}
g = 0.078 + 1.88/[J(J+1)].
\end{equation}

For the $5_1-6_0$~A$^+$ 6.7~GHz methanol transition we find that the
Zeeman splitting is $\Delta V_Z{\rm [km s^{-1}]} = 0.049 B{\rm
  [G]}$. Thus, we can place an upper limit to the magnetic field in
the 6.7~GHz methanol broadline maser region in W3(OH) of $22$~mG. This
is consistent with the magnetic field strength of $B=2$--$11$~mG found
using the ground-state OH masers at 1612, 1665 and 1667~MHz
\citep[e.g][]{Wright04b}, noting that the ground-state and
  6.8~GHz methanol masers are excited under similar physical
  conditions. It is also consistent with the magnetic field of $\sim
15$~mG found from the 6~GHz OH masers \citep{Desmurs98, Etoka05}
and the magnetic field of $\sim 10$~mG found from the 13.44~GHz OH
masers \citep{Baudry98}, both of which were detected in the broadline
region, coincident with the 6.7~GHz methanol masers.

\section{Conclusion}

MERLIN has been used to produce the first maps of the polarization of
6.7~GHz methanol masers in the star-forming region W3(OH). Linear
polarization of up to $\sim 8$~per~cent, similar in level to that
found from the 12.2~GHz methanol masers at the same velocities, has
been found. We determined an upper limit of $22$~mG to the magnetic
field strength through the RHC-LHC cross-correlation method. As the
6.7~GHz methanol masers are much less affected by internal and
external Faraday rotation than the lower frequency OH masers, they are
excellent probes of the magnetic field direction in the maser
region. After correction for Faraday rotation we find that the magnetic
field is orientated parallel to the extended N-S filamentary structure
detected in methanol and OH maser observations.

\section*{Acknowledgments}

WV acknowledges support by a Marie Curie Intra-European Fellowship within the 6th European Community Framework Program under contract number MEIF-CT-2005-010393



\begin{thebibliography}{}

\bibitem[\protect\citeauthoryear{{Bartkiewicz}, {Szymczak}, {Cohen} \&
  {Richards}}{{Bartkiewicz} et~al.}{2005}]{Bartkiewicz05}
{Bartkiewicz} A.,  {Szymczak} M.,  {Cohen} R.~J.,    {Richards} A.~M.~S.,
  2005, \mnras, 361, 623

\bibitem[\protect\citeauthoryear{{Baudry} \& {Diamond}}{{Baudry} \&
  {Diamond}}{1998}]{Baudry98}
{Baudry} A.,  {Diamond} P.~J.,  1998, \aap, 331, 697

\bibitem[\protect\citeauthoryear{{Cragg}, {Sobolev} \& {Godfrey}}{{Cragg} et~al.}{2002}]{Cragg02}
{Cragg} D.~M.,  {Sobolev} A.~M.,  {Godfrey} P.~D., 2002, \mnras, 331, 521

\bibitem[\protect\citeauthoryear{{Desmurs}, {Baudry}, {Wilson}, {Cohen} \&
  {Tofani}}{{Desmurs} et~al.}{1998}]{Desmurs98}
{Desmurs} J.~F.,  {Baudry} A.,  {Wilson} T.~L.,  {Cohen} R.~J.,    {Tofani} G.,
   1998, \aap, 334, 1085

\bibitem[\protect\citeauthoryear{{Dodson}, {Ohja} \& {Ellingsen}}{{Dodson} et al.}{2004}]{Dodson04}
{Dodson} R.,  {Ojha} R.,  {Ellingsen} S.~P., 2004, \mnras, 351, 779

\bibitem[\protect\citeauthoryear{{Dreher} \& {Welch}}{{Dreher} \&
  {Welch}}{1981}]{Dreher81}
{Dreher} J.~W.,  {Welch} W.~J.,  1981, \apj, 245, 857

\bibitem[\protect\citeauthoryear{{Ellingsen}}{{Ellingsen}}{2002}]{Ellingsen02}
{Ellingsen} S.~P.,  2002, IAU Symposium, 206, 151

\bibitem[\protect\citeauthoryear{{Etoka}, {Cohen} \& {Gray}}{{Etoka}
  et~al.}{2005}]{Etoka05}
{Etoka} S.,  {Cohen} R.~J.,    {Gray} M.~D.,  2005, \mnras, 360, 1162

\bibitem[\protect\citeauthoryear{{Fish} \& {Reid}}{{Fish} \&
  {Reid}}{2006}]{Fish06}
{Fish} V.~L.,  {Reid} M.~J.,  2006, \apjs, 164, 99

\bibitem[\protect\citeauthoryear{{Goldreich}, {Keeley} \& {Kwan}}{{Goldreich}
  et~al.}{1973}]{Goldreich73}
{Goldreich} P.,  {Keeley} D.~A.,    {Kwan} J.~Y.,  1973, \apj, 179, 111

\bibitem[\protect\citeauthoryear{{Hartquist}, {Menten}, {Lepp} \&
  {Dalgarno}}{{Hartquist} et~al.}{1995}]{Hartquist95}
{Hartquist} T.~W.,  {Menten} K.~M.,  {Lepp} S.,    {Dalgarno} A.,  1995,
  \mnras, 272, 184

\bibitem[\protect\citeauthoryear{{Harvey-Smith} \& {Cohen}}{{Harvey-Smith} \& 
{Cohen}}{2006}]{HarveySmith06}
{Harvey-Smith} L.,  {Cohen} R.J., 2006, \mnras, accepted (HSC06)

\bibitem[\protect\citeauthoryear{{Imai}, {Horiuchi}, {Deguchi} \&
  {Kameya}}{{Imai} et~al.}{2003}]{Imai03}
{Imai} H.,  {Horiuchi} S.,  {Deguchi} S.,    {Kameya} O.,  2003, \apj, 595, 285

\bibitem[\protect\citeauthoryear{{Jen}}{{Jen}}{1951}]{Jen51}
{Jen} C.~K.,  1951, Physical Review, 81, 197

\bibitem[\protect\citeauthoryear{{Koo}, {Williams}, {Heiles} \& {Backer}}{{Koo}
  et~al.}{1988}]{Koo88}
{Koo} B.-C.,  {Williams} D.~R.~D.,  {Heiles} C.,    {Backer} D.~C.,  1988,
  \apj, 326, 931

\bibitem[\protect\citeauthoryear{{Menten}}{{Menten}}{1991}]{Menten91}
{Menten} K.~M.,  1991, \apjl, 380, L75

\bibitem[\protect\citeauthoryear{{Modjaz}, {Moran}, {Kondratko} \&
  {Greenhill}}{{Modjaz} et~al.}{2005}]{Modjaz05}
{Modjaz} M.,  {Moran} J.~M.,  {Kondratko} P.~T.,    {Greenhill} L.~J.,  2005,
  \apj, 626, 104

\bibitem[\protect\citeauthoryear{{Sarma}, {Troland} \& {Romney}}{{Sarma}
  et~al.}{2001}]{Sarma01}
{Sarma} A.~P.,  {Troland} T.~H.,    {Romney} J.~D.,  2001, \apjl, 554, L217

\bibitem[\protect\citeauthoryear{{Turner} \& {Welch}}{{Turner} \&
  {Welch}}{1984}]{Turner84}
{Turner} J.~L.,  {Welch} W.~J.,  1984, \apjl, 287, L81

\bibitem[\protect\citeauthoryear{{Vlemmings}, {Diamond}, {van Langevelde} \&
  {Torrelles}}{{Vlemmings} et~al.}{2006}]{Vlemmings06}
{Vlemmings} W.~H.~T.,  {Diamond} P.~J.,  {van Langevelde} H.~J.,    {Torrelles}
  J.~M.,  2006, \aap, 448, 597

\bibitem[\protect\citeauthoryear{{Wright}, {Gray} \& {Diamond}}{{Wright}
  et~al.}{2004a}]{Wright04a}
{Wright} M.~M.,  {Gray} M.~D.,    {Diamond} P.~J.,  2004a, \mnras, 350, 1253

\bibitem[\protect\citeauthoryear{{Wright}, {Gray} \& {Diamond}}{{Wright}
  et~al.}{2004b}]{Wright04b}
{Wright} M.~M.,  {Gray} M.~D.,    {Diamond} P.~J.,  2004b, \mnras, 350, 1272

\bibitem[\protect\citeauthoryear{{Xu}, {Reid}, {Zheng} \& {Menten}}{{Xu}
  et~al.}{2006}]{Xu06}
{Xu} Y.,  {Reid} M.~J.,  {Zheng} X.~W.,    {Menten} K.~M.,  2006, Science, 311,
  54

\end{thebibliography}

\label{lastpage}

\end{document}